# Analysis of Trace Impurities in Semiconductor Gas via Cavity-Enhanced Direct Frequency Comb Spectroscopy

K.C. Cossel[1*], F. Adler[1], K.A. Bertness[2], M.J. Thorpe[1], J. Feng[3], M.W. Raynor[3], J. Ye[1]

[1]JILA, National Institute of Standards and Technology and University of Colorado

Department of Physics, University of Colorado, Boulder, Colorado 80309-0440, USA

[2]National Institute of Standards and Technology, Boulder, Colorado 80305-3328, USA

[3]Matheson Tri-Gas, Longmont, Colorado 80501, USA

*kevin.cossel@colorado.edu

**Cavity-enhanced direct frequency comb spectroscopy (CE-DFCS) has demonstrated powerful potential for trace gas detection based on its unique combination of high bandwidth, rapid data acquisition, high sensitivity, and high resolution, which is unavailable with conventional systems. However, previous demonstrations have been limited to proof-of-principle experiments or studies of fundamental laboratory science. Here we present the development of CE-DFCS towards an industrial application – measuring impurities in arsine, an important process gas used in III-V semiconductor compound manufacturing. A strongly absorbing background gas with an extremely complex, congested, and broadband spectrum renders trace detection exceptionally difficult, but the capabilities of CE-DFCS overcome this challenge and make it possible to identify and quantify multiple spectral lines associated with water impurities. Further, frequency combs allow easy access to new spectral regions via efficient nonlinear optical processes. Here, we demonstrate detection of multiple potential impurities across 1.75-1.95 μm (5710-5130 cm$^{-1}$), with a single-channel detection sensitivity of**

~4 x $10^{-8}$ cm$^{-1}$ Hz$^{-1/2}$ in nitrogen and specifically, absorption sensitivity of ~4 x $10^{-7}$ cm$^{-1}$ Hz$^{-1/2}$ for trace water doped in arsine.

## *1. Introduction*

Arsine (AsH$_3$) and phosphine (PH$_3$) are important process gases used in the production of III-V semiconductors via metal organic chemical vapor deposition (MOCVD) [1-4]. These compounds are used in devices ranging from high-brightness light emitting diodes and high power laser diodes to solar cells. Trace levels of contaminants present in the process gases can result in unintentional doping and lattice defects. The inclusion of these dopants gives rise to additional energy levels in the bandgap of the host semiconductor, leading to undesired changes of its electrical and optical properties. In particular, oxygen incorporation in III-V semiconductors has been shown to form a deep recombination level, resulting in a decrease of photoluminescence efficiency and carrier lifetimes as well as reduced device reliability. A primary source of oxygen impurities is the presence of traces of water vapor in the precursor gases, which has been shown to negatively impact the semiconductor at the level of 10-100 parts per billion (ppb) [4]. Because of its ubiquity and low vapor pressure, water is extremely difficult for the manufacturer to completely remove. Furthermore, water impurities can be introduced from contamination in the transfer lines at the point of use; therefore, on-line monitoring of water concentrations during semiconductor growth is desirable. In addition to water vapor, many other impurities including carbon dioxide, hydrocarbons (methane and ethane), hydrogen sulfide, silane and germane must be controlled in the process gas [2].

Current systems, employing different techniques, capable of detecting water at the sub-100-ppb level all have significant drawbacks [2]. For example, Fourier transform infrared (FTIR) [5] spectrometers require long acquisition times owing to the low spectral brightness of their thermal light source. In addition, the required sensitivity can only be achieved by using a highly optimized research system with a long pathlength gas cell, a high sensitivity detector, and an ultrahigh purity nitrogen purge system. Another potential technique is negative ion-atmospheric pressure ionization mass spectrometry (APIMS) [6], but these expensive systems are large and complex, not suitable for on-line monitoring and are mainly limited to inert background gases. Current research has focused on laser-based spectroscopy systems such as tunable diode laser spectroscopy (TDLAS) or cavity ring-down spectroscopy (CRDS) [1, 3, 7], which provide highly sensitive measurements with rapid acquisition times; however, these systems typically target one or two absorption lines of the desired impurity species. Thus, multiple contaminants cannot be easily identified and quantified, while unexpected impurities can significantly degrade the reliability of the measurement. In a recent study [4], it was necessary to use gas chromatography with mass spectrometry, atomic emission spectroscopy, and pulsed discharge ionization detection in addition to CRDS, to check for all of the critical impurities in arsine. Cavity-enhanced direct frequency comb spectroscopy (CE-DFCS) [8, 9] provides both high sensitivity and broad bandwidth, which enabled multi-species trace detection in breath near 1.5 µm [10]. Previous cw-CRDS experiments measuring water in phosphine at 940 nm [7] and 1400 nm [3] were limited by competing background gas absorption and by the line strength of the water transition. The 1.75–1.95 µm (5710–5130 cm$^{-1}$) region contains bands of several important impurities, including a water band at 1.85 µm (5400

cm$^{-1}$) that is stronger than the 1400 nm band by a factor of 2-3, and should be a somewhat transparent region of the arsine absorption spectrum [11]. Therefore, it is a compelling range to explore with CE-DFCS.

CE-DFCS is a recently developed spectroscopic technique that potentially overcomes the difficulty of slow speed with continuous-wave (cw) laser based spectroscopy [9, 10], while maintaining the high-resolution capability. In addition to high detection sensitivity obtained from the use of an optical enhancement cavity (as in CRDS), CE-DFCS has broad bandwidth available from the femtosecond optical frequency comb. Furthermore, the use of ultrashort pulses enables highly efficient nonlinear frequency conversion, further increasing the possible bandwidth and providing easy access to spectral regions that are difficult to access with conventional methods. We use this latter advantage in this paper, demonstrating for the first time direct frequency comb spectroscopy in a region accessed by supercontinuum generation – in this case, 1.75–1.95 µm (5710–5130 cm$^{-1}$). This spectral region is mostly unexplored via laser-based spectroscopy owing to the lack of widely tunable lasers operating here. In fact, DFB diode lasers have only recently become available in this spectral region [12], but still with limited tuning range. Two possible broadly tunable sources are optical parametric oscillators [13] or difference frequency generation [14]; however, these cw systems must be carefully scanned to cover the full spectral region. Our CE-DFCS approach provides high sensitivities across more than 2000 detection channels distributed over a broad, simultaneous bandwidth in a robust and compact system. Furthermore, we have performed CE-DFCS for the first time with focus on an industrial application (i.e., trace detection in a strongly absorbing process gas), where the bandwidth is critical for distinguishing impurity signals from background absorption, along with high resolution for making unambiguous identifications.

## *2. Experimental Details*

### 2.1 Frequency Comb Source

We used a home-built mode-locked Erbium-doped fiber ring laser [15] that provides 130 mW average power at a repetition rate of approximately 250 MHz. The laser output was then amplified with a single mode Er:fiber amplifier to produce 81-fs pulses with 400 mW average power. These parameters are readily achievable with commercially available fiber lasers as well. After the amplifier and polarization control optics, all of the light was coupled into 10 cm of standard single mode fiber spliced to a 6 cm long piece of highly non-linear silica fiber (OFS Specialty Photonics) to provide spectral broadening. The spectrum (covering 1.2-2.1 µm or 8300-4700 $cm^{-1}$) from this fiber is shown in Fig. 1 as optimized for generation of a frequency comb near 1.85 µm; however, it is easy to change the spectrum by varying polarization and input pulse chirp. The average power after a 40-nm bandwidth filter centered at 1.86 µm was 17 mW. Because the laser and nonlinear spectrum generation were both fiber-based, this entire system was compact and robust – requiring almost no adjustment from day to day.

### 2.2 Optical Cavity and Sample Cell

The optical resonator used for enhancing the absorption detection was a linear Fabry-Pérot cavity with one 2-m radius concave mirror and one flat mirror and a peak finesse of 30,000 (mirror reflectivity ~ 0.9999). In order to obtain accurate absorption values, we characterized the mirror reflectivity via wavelength resolved ringdown measurements. For this we recorded ringdown traces of an empty (under vacuum) cavity in 10 nm wavelength increments using a monochromator and a fast extended-InGaAs photodiode. Several measurements for each wavelength were averaged, and the resulting data was fit with an 8$^{th}$ order polynomial.

The cavity length (~60 cm) was adjusted so that the free spectral range was matched to the comb repetition rate. In addition, the comb-offset frequency ($f_0$) was adjusted via the fiber laser pump power to optimize the transmission of the comb through the enhancement cavity [9]. The useful spectral bandwidth of the cavity was about 200 nm (600 cm$^{-1}$); however, owing to the varying free spectral range frequency due to cavity dispersion, it is not possible to simultaneously match all comb modes and cavity modes over this bandwidth. To overcome this limitation and to synchronize the cavity and comb during measurements, the comb modes were dithered (by changing the laser cavity length and therefore the repetition rate) using a triangle waveform with an amplitude of 150 kHz at a frequency of 7.5 kHz around the cavity modes, and slow feedback to the enhancement cavity length (via a piezo) was used to keep the time gap between the successive transmission peaks constant [9]. Since our integration time per individual record was around 150 ms, we integrated over multiple dither cycles. This dither process effectively allowed us to couple the comb to the cavity over the full spectral bandwidth of the mirrors. Furthermore, if the comb were locked to the cavity without dither, the transmitted intensity noise would be significantly increased due to FM-AM conversion from cavity vibrations and laser frequency jitter. Dithering the comb reduces this noise and also simplifies locking.

The sample gases flowed through the optical buildup cavity for detection. To measure and quantify impurities in arsine, our gas handling system was designed to add trace contaminants to arsine at well-controlled mixing ratios. For this purpose we obtained a calibrated mixture of 10 parts per million (ppm) each of $CO_2$, $CH_4$, and $H_2S$ in a nitrogen gas cylinder. In addition, we added small amounts of water vapor via a diffusion vial. Typically, the total pressure in the sample cell was ∼ 200 torr. The gas system was designed to be very flexible for this experiment; however, it was not optimized for switching speed due to the small diameter of the installed tubing.

## 2.3 Virtually Imaged Phase Array (VIPA) Spectrometer

The light transmitted from the cavity was analyzed with a two-dimensional dispersive spectrometer based on a virtually imaged phased array (VIPA), a cross-dispersion grating, and a 2D camera [10, 16-18]. This spectrometer system provides high resolution (900 MHz or 0.031 $cm^{-1}$) in one dimension, while maintaining broad bandwidth (∼20 nm or 50 $cm^{-1}$, limited by the size of the camera's imaging sensor) in the orthogonal dimension. Light was coupled into the VIPA etalon (51 GHz free spectral range) with a horizontal 6 cm focal length cylindrical lens. The etalon was tilted approximately 2° from vertical to provide a high dispersion in the vertical direction. A cross-dispersing grating was used to separate the mode orders in the horizontal direction. Finally, the light was imaged onto the 320x256-pixel InSb focal plane array camera with a 25 cm focal length lens, resulting in images as shown in Fig. 1. Each bright stripe corresponds to a mode order of the etalon resolved with the grating; 900 MHz (0.03 $cm^{-1}$) resolution is obtained in the vertical direction due to the high dispersion of the VIPA, while the cross dispersion allows us to collect over 2000 spectral channels simultaneously. Images were collected alternating between sample and reference gases, which provides a differential measurement of the change in transmitted power with and without absorber. Typically, we average 20 images (at ~ 150 ms integration time per image) with sample gas; then switched to reference gas and averaged another 20 images. This sequence was repeated to average down to the desired sensitivity.

## 2.4 Spectral Recovery

The two final sets of images, one with sample gas in the cavity and one with nitrogen reference gas, provided a measurement of $\Delta I/I_0$ for each detection channel, i.e., the fractional change in integrated power due to absorption in the cavity. This quantity is equivalent to measuring the fractional change in cavity ringdown time ($\Delta\tau/\tau_0$) because the integrated intensity of an exponential decay is proportional to $\tau$ and then the ratio removes the proportionality constant.

The quantity $\Delta\tau/\tau$ can be expressed as a function of standard absorption per centimeter ($\alpha$) by using

$$\Delta\tau/\tau_0 = (\tau(\alpha) - \tau_0)/\tau_0$$

$$\tau(\alpha) = \frac{2L}{c(1 - R^2 e^{-2\alpha L})} \; ; \; \tau_0 = \frac{2L}{c(1 - R^2)}$$

where $L$ is the cavity length (obtained by counting the laser repetition rate) and $R$ is the single mirror reflectivity [19]. This can be solved for $\alpha$ to give

$$\alpha = -\frac{1}{2L}\ln\left[\frac{1}{R^2}\left(1-\frac{1-R^2}{1-\frac{\Delta I}{I_0}}\right)\right],$$

where $\Delta I/I_0$ has been substituted for $\Delta\tau/\tau_0$. This equation was used to evaluate the per channel absorption to give the final spectrum. The frequency axis was calibrated using known line positions from the HITRAN database [20].

## *3. Results*

### 3.1 Trace gasses in $N_2$

To calibrate our CE-DFCS system, we first recorded the spectrum of the impurity gas cylinder plus 2.5 ppm $H_2O$ in nitrogen. The measured absorption spectra are plotted below the axis as inverted peaks in black in Fig. 2.[1] For comparison, known spectral lines from HITRAN [20] (for $CO_2$, $H_2O$, and $CH_4$) and PNNL [11] (for $H_2S$) are plotted above the axis. This overall spectrum is a composite of 11 individual spectra, spanning over 700 $cm^{-1}$ (200 nm) spectral bandwidth with a step size of 0.014 $cm^{-1}$. The total acquisition time was approximately 10 hours; however, this was severely limited by the switching speed of the gas flow system. By using a gas handling system designed for rapid gas switching (large tube diameter, short lines, high flow rates, etc.) switching times could be significantly reduced, resulting in an estimated acquisition time of ~15 minutes per individual spectrum (~2.5 hours for the full bandwidth) to achieve the current sensitivity. In addition, a carefully designed system using two separate cavities (one for sample gas and one for reference gas) can reduce common-mode amplitude noise and would require no gas switching, thus improving sensitivity and further reducing the required acquisition time.

---

[1] A detailed spectrum is available at:
http://jila.colorado.edu/yelabs/pubs/scienceArticles/2010/ImpurityCylinderSpectrum.pdf

The high resolution of the system can be observed clearly in Figs. 2a-c, where three separate regions of the spectrum are expanded, each with a 30-cm$^{-1}$ span. Fig. 2a shows a reference FTIR spectrum ($H_2S$, in purple, from PNNL [10]), reference line positions for $CO_2$ (green, from HITRAN), and a spectrum with 1.1 GHz (0.035 cm$^{-1}$) resolution obtained with broadband CE-DFCS (black line). The current resolution is ideal for the 1-2 GHz wide Doppler and pressure broadened lines observed here. Since 5-10 comb modes sample each absorption feature, lineshape distortion is not an issue and single comb mode resolution is not required; however, if desired, resolutions of below 250 kHz ($8\times10^{-6}$ cm$^{-1}$) are obtainable with modifications to resolve single comb lines [9]. In addition, Figs. 2a-c illustrate that the relative frequency accuracy of the spectrum is high; we estimate it to be better than 100 MHz (0.0033 cm$^{-1}$) across the full 700 cm$^{-1}$ spectral bandwidth by qualitatively comparing measured line positions with HITRAN [20]. Again, by resolving individual comb lines, absolute frequencies with sub-kHz accuracies are possible. Indeed, CE-DFCS can be an effective and accurate tool for future additions and modifications to molecular absorption databases.

Since we are able to observe many absorption lines for each species, the concentration and minimum detectable absorption may be determined with a modification of the Hubaux-Vos regression method [21], which is similar to methods used by the International Union of Pure and Applied Chemistry [22] and Semiconductor Equipment and Materials International [23], using a single measured spectrum at a fixed concentration. All of these methods use a calibration curve of measured absorption versus sample concentration to determine the range of noise at zero concentration. Instead of measuring a single absorption feature as a function of sample concentration, we determine the measured absorption as a function of predicted absorption for lines with different strengths. This result allows us to determine both the noise-equivalent absorption and the concentration of the sample. In other words, one spectrum allows us to collect noise statistics over a wide range of signal strengths. To do this, we first predict the spectrum by approximating the concentration and using available HITRAN [20] data with modifications due to pressure broadening, Doppler broadening, spectrometer resolution, and digital filters. We then plot each observed peak height versus the predicted height of the corresponding peak, as shown in Fig. 3a for $CO_2$ and Fig. 3b for $H_2O$ (note that saturation was observed at higher absorptions, these points were not included in the analysis). The slope of a linear fit to this comparison data gives a correction factor from the approximated concentration to the true concentration and the uncertainty in the slope provides the uncertainty in the true concentration. In addition, the minimum detectable absorption is given by the $3\sigma$ standard error of the intercept, which specifies the measured absorption value that can be considered a non-zero true absorption with 99.86% confidence. This is shown graphically in Fig. 3 by the value of the upper confidence interval when it intersects the y-axis. Since the confidence intervals shown are for the fit, it is not expected that all data should lie within them; the standard error of the fit is much smaller than the standard deviation of each data point due to the large number of points. The error on the intercept is roughly the same as the standard deviation for each data point.

For $CO_2$ in the nitrogen gas, we measure a concentration of 9.7 ± 0.2 ppm (3σ uncertainty in slope of fit), consistent with the factory-specified impurity level in the cylinder at 10 ppm ± 10%. In addition, we obtain a minimum detectable absorption (3σ) of $4.5\times10^{-9}$ $cm^{-1}$, corresponding to 325 ppb minimum detectable concentration by using the line strength of the strongest measured line. This data was recorded with a total integration time (T) of 180 s (time includes reference and sample images, but not gas switching times), which gives a 1-Hz minimum detectable absorption of $4.2\times10^{-8}$ $cm^{-1}$ $Hz^{-1/2}$. For "boxcar" style integration as we do here, the bandwidth is defined as 2/T, analogous to the bandwidth when calculating shot noise. Similarly for $H_2O$, we measure 2.50 ± 0.12 ppm, with an absorption sensitivity of $5.5\times10^{-9}$ $cm^{-1}$ or a minimum detectable concentration of 7 ppb. Since this absorption sensitivity is uniform across our measured spectral range, we project a minimum detectable concentration of 700 ppb for $CH_4$ and 370 ppb for $H_2S$. We also note the advantage of a dramatically increased dynamic range (larger than the single concentration data in Fig. 3) obtained from simultaneously studying lines with significantly different strengths. Even though strong absorption peaks (above $2\times10^{-7}$ $cm^{-1}$) provide signals that are saturated at the measured concentration, at lower concentrations they will be in a linear regime while the weak lines will be too small to measure. Additionally, lines that are too weak to measure at the current concentration will be detectable at higher concentrations. By using the range of water line strengths given in HITRAN [20] for this wavelength region, we estimate a measurement range for water concentration from 7 ppb to 100 ppm with this system.

## 3.2 Impurities in arsine

After calibrating the system, we filled the high finesse optical cavity with arsine interspersed with impurities. The same mixture measured in nitrogen was added to a purified arsine gas in a ratio of 1:8. Figure 4 shows the composite spectrum of trace water recorded in the background of arsine gas at ~160 torr.[2] Again, measured absorption features are plotted as inverted peaks below the axis, with available HITRAN [20] spectral lines on water plotted above the axis. It is apparent that the absorption of arsine continues to increase rapidly towards both ends of the spectral window; past the plotted range, the arsine absorption begins to decrease at ~1.94 µm, but even at 1.97 µm it was still too strong to collect reliable spectra of trace species. The density of the arsine absorption obscured the other impurity lines, making it only possible to observe water in this experiment. Because of the strongly absorbing background gas, it is clearly a major advantage to have a spectroscopy system based on CE-DFCS providing broad bandwidth and high spectral resolution. Such a system allows us to easily identify and quantify water lines that are isolated from arsine absorption features, as seen by the two zoomed-in bottom panels of Fig. 4. From this data (see Fig. 3c) we determine a water concentration of $1.27 \pm 0.08$ ppm and an absorption sensitivity of $2.4 \times 10^{-8}$ cm$^{-1}$ with an integration time of 600 s, which corresponds to a minimum detectable concentration for water in arsine of 31 ppb.

---

[2]   A detailed spectrum is available at:
   http://jila.colorado.edu/yelabs/pubs/scienceArticles/2010/ArsineWithWater.pdf

## *4. Conclusions*

This first demonstration of CE-DFCS for impurity monitoring of industrial process gases has shown an absorption detection sensitivity (3σ) of $3.9\times10^{-7}$ cm$^{-1}$ Hz$^{-1/2}$ for water in arsine around 1.85 μm (5400 cm$^{-1}$). The 200 nm bandwidth of our system also allowed us to look at absorption regions of other impurities ($CH_4$, $CO_2$, and $H_2S$), which were obscured by densely spaced arsine absorption lines. However, the same system can be easily extended anywhere from 1.2–2.1 μm (8300-4700 cm$^{-1}$) with additional mirrors or potentially with a single cavity using Brewster angle prisms [24] to cover additional impurity bands. For future real-time monitoring in industrial conditions acquisition times and detection sensitivities should be improved. As discussed previously, the current acquisition time was limited by the gas handling system and could be significantly reduced. A two-cavity system or the recently demonstrated two-comb multi-heterodyne technique [25-28] will further reduce intensity noise by decreasing the time between sample and reference spectra. Additionally, extending spectral coverage of combs into the mid-IR [29] provides a promising direction for improvement in detection sensitivity and searching for new transparency windows of industrial process gasses. Overall, the capabilities of comb spectroscopy for real-world applications provide a clear path towards further improvements and application-oriented developments for this technique.

## Acknowledgements

We dedicate this paper to Dr. Jun Feng (1957 - 2010), who initiated this collaborative project. We would like to thank R. Holzwarth and T. Wilken for helpful discussions. Funding was provided by AFOSR, DARPA, DTRA, Agilent, and NIST. K.C.C. acknowledges support from the NSF

Graduate Research Fellowship. F.A. is partially supported by the Alexander von Humboldt Foundation.

**Figure 1**: **Overview of experiment.** *A mode-locked Er:fiber laser is spectrally broadened using a highly non-linear fiber (HNLF) to provide comb light from 1.2 – 2.1 µm, which is then coupled into a high-finesse Fabry-Pérot cavity (F ~ 30,000). A small amount of the cavity transmission is used to match the laser repetition rate and the cavity free spectral range. The rest of the cavity transmission is dispersed using a 2D VIPA spectrometer and then imaged onto an InSb focal plane array. Raw images with either nitrogen (reference gas) or arsine ($AsH_3$, or other sample gas) in the cavity are converted into a 2D absorption image (1 – sample/reference) and then into a 1D spectrum. A single-shot spectrum as shown covers over 20 nm.*

**Figure 2: Trace gases in nitrogen**. *The spectrum of 10 ppm methane ($CH_4$), 10 ppm carbon dioxide ($CO_2$), 10 ppm hydrogen sulfide ($H_2S$), and 2.5 ppm water ($H_2O$) in nitrogen gas. Measured absorption peaks are shown in black below the axis. For comparison, spectral lines for $CH_4$, $CO_2$, and $H_2O$ (line strength divided by 10) available from the HITRAN database as well as $H_2S$ lines from the PNNL database are plotted above the central axis. Expanded spectral windows a, b, and c illustrate high-resolution absorption spectra of $CO_2$ and $H_2S$, $H_2O$ and $H_2S$, and $H_2O$ and $CH_4$, respectively.*

**Figure 3: Determination of absorption sensitivities.** *(color online) Double logarithmic plots of measured peak absorption versus predicted peak absorption from the HITRAN database for 10 ppm $CO_2$ in $N_2$ (a), 2.5 ppm $H_2O$ in $N_2$ (b), and 1 ppm $H_2O$ in arsine (c). Linear fits (dashed) of the data sets were performed up to a measured absorption of $2 \times 10^{-7}$, where signal saturation*

*becomes noticeable. The corresponding 3σ (99.87%) confidence intervals (solid lines) are plotted along with the linear fits. Note that these confidence intervals are for the fits to the entire data. The 3σ error on the intercept gives the minimum detectable absorption, the slope of the fit gives a correction factor for the concentration, and the error on the slope gives the uncertainty in the concentration.*

**Figure 4: Trace water in arsine.** *The measured spectrum of 1.27 ppm of water vapor in arsine gas is plotted in black below the axis and the HITRAN reference lines for water are shown in blue above the axis. The arsine absorption continues to increase both above 5650 cm$^{-1}$ and below 5350 cm$^{-1}$, precluding any additional measurements. The two insets show easily resolvable water lines even in a background absorbing gas.*

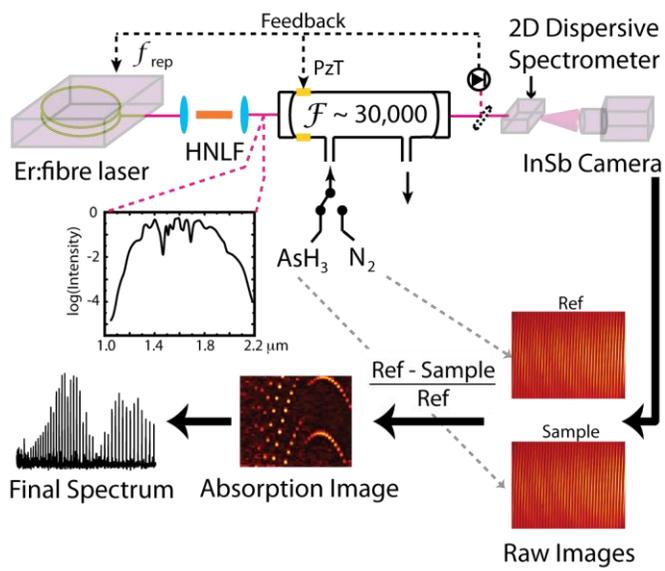

**Figure 1**

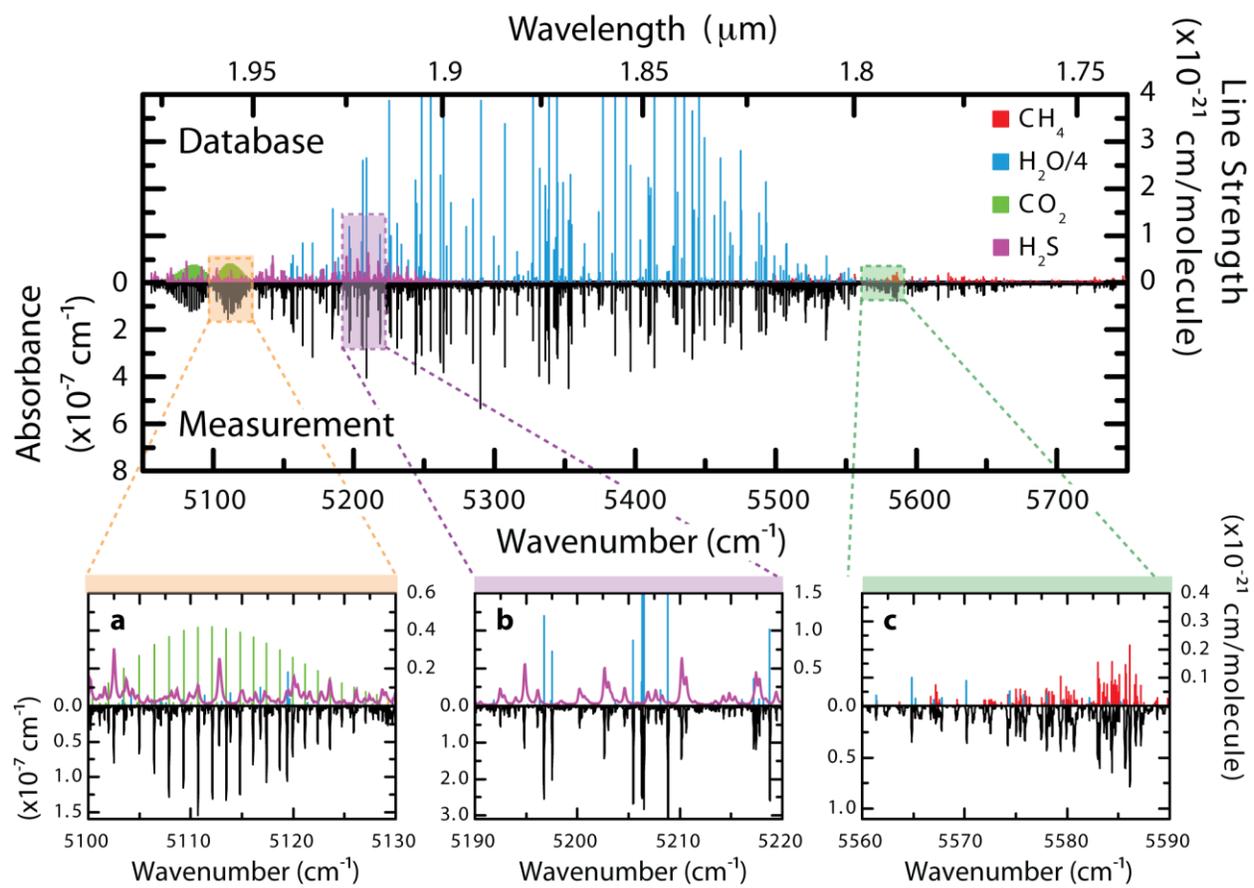

**Figure 2**

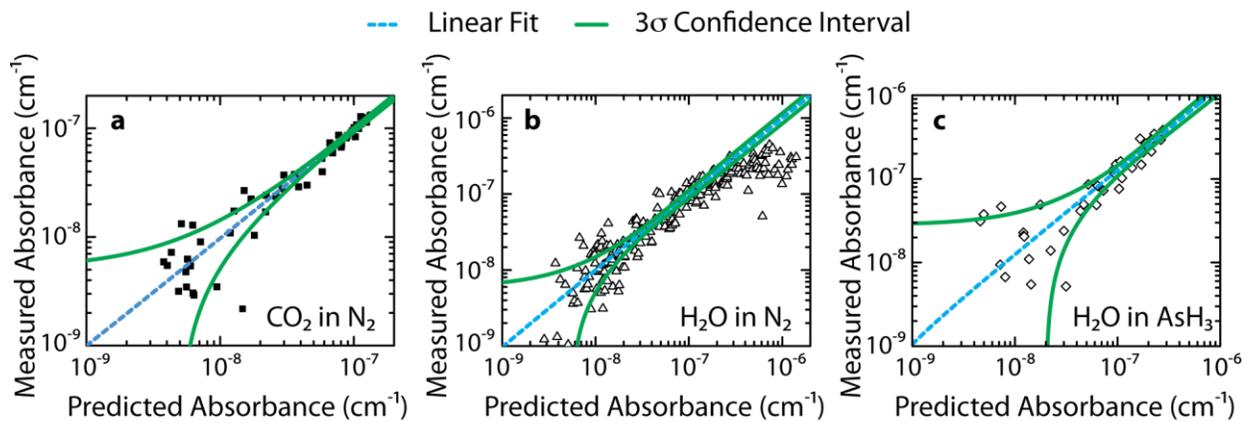

**Figure 3**

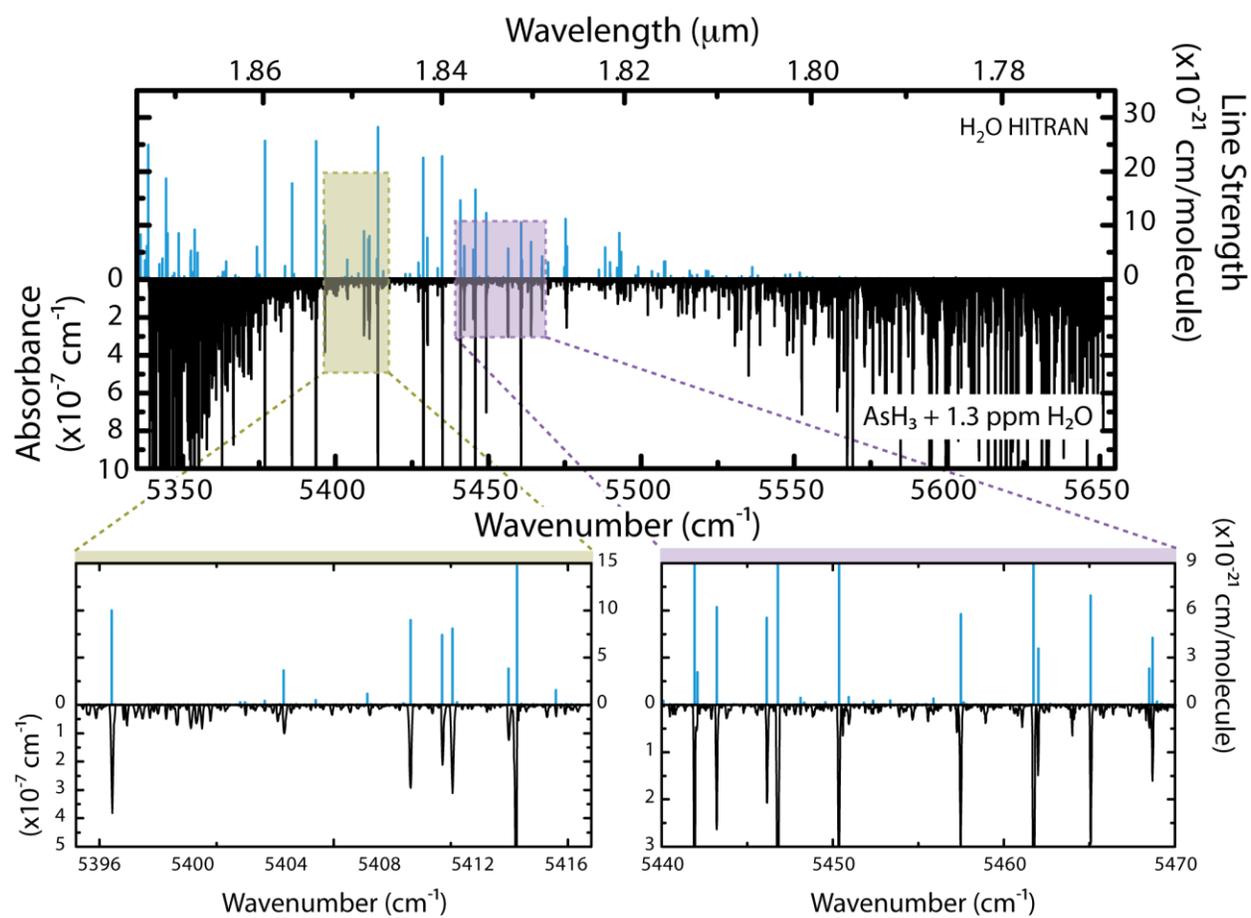

**Figure 4**